\documentclass[preprint,aps,showpacs]{revtex4}
\usepackage{graphicx}

\def\be7pg{$^7Be(p,\gamma)^8B$}
\def\xbe7{$^7Be$}
\def\b8{$^8B$}
\def\S17{$S_{17}(0)$}
\def\xs17{$S_{17}$}
\def\s34{$S_{34}(0)$}
\def\xpm{$\pm$}
\def\xchi{$\chi ^2 / \nu$}

\begin{document}

\title{Is There a Significant Difference Between the Results of the Coulomb Dissociation 
of \b8 and the Direct Capture \be7pg Reaction?}
\thanks{Work Supported by USDOE Grant No. DE-FG02-94ER40870.}

\author{Moshe Gai}
\affiliation{Laboratory for Nuclear Science at Avery Point, 
University of Connecticut, 1084 Shennecossett Rd, Groton, CT 06340-6097.\\
and\\
Department of Physics, WNSL Rm 102, Yale University, PO Box 208124, 
272 Whitney Avenue, 
New Haven, CT 06520-8124.
\    \\
e-mail: moshe.gai@yale.edu, URL: http://www.phys.uconn.edu}

\begin{abstract}

Recent claims of the Seattle group of evidence of "slope difference between 
CD [Coulomb Dissociation] and direct [capture] results" are based on 
wrong and selective data. When the RIKEN2 data are included correctly, and previously 
published Direct Capture (DC) data are also included, we observe only a 1.9 sigma 
difference in the extracted so called "scale independent slope (b)", considerably smaller than 
claimed by the Seattle group. The very parameterization used by the Seattle group 
to extract the so called b-slope parameter has no physical 
foundation. Considering the physical slope (S' = dS/dE) above 300 keV, 
we observe a 1.0 sigma agreement between slopes (S') 
measured in CD and DC, refuting the need for new theoretical investigation.
The claim that \S17 values extracted from 
CD data are approximately 10\% lower than DC results, is based on
misunderstanding of the CD method. Considering all of the published CD \S17 
results, with adding back an unconfirmed E2 correction of the MSU data, yields  
very consistent \S17 results that agree with recent DC measurements of 
the Seattle and Weizmann groups.  The recent correction of the b-slope parameter 
(0.25 $MeV^{-1}$) suggested by Esbensen, Bertsch and Snover was applied to the 
wrong b-slope parameter calculated by the Seattle group. When considering the correct 
slope of the RIKEN2 data, this correction in fact leads to a very small b-slope parameter 
(0.14 $MeV^{-1}$), less than half the central value 
observed for DC data, refuting the need to correct the RIKEN2 data.
In particular it confirms that the E2 contribution in the RIKEN2 data is negligible.
The dispersion of measured \S17 is mostly due to disagreement  among 
individual DC experiments and not due to either experimental or theoretical 
aspects of CD. Additional uncertainty exists due to theoretical extrapolation 
procedure that for example was recently estimated by Descouvemont to be 6\%, more 
than a factor of two larger than suggested by the Seattle group. Uncertainty of 
the slope (S'), that thus far was not measured with high precision, leads to a substantial 
error of extrapolated \S17.

\end{abstract}


\preprint{UConn-40870-0XXX}

\maketitle

\section{Introduction}

The method of Coulomb Dissociation (CD) was developed in the pioneering 
work of Baur, Bertulani and Rebel \cite{Baur} and has been applied to the case 
of the CD of \b8 \cite{Mot94,Kik,Iw99,Sch03}. These data were analyzed with a 
remarkable success using only first order Coulomb interaction that includes only 
E1 contribution. Indeed early attempts to refute this analysis by introducing a 
non-negligible E2 contribution were shown \cite{Gai} to arise from a misunderstanding 
of the RIKEN1 data; hereafter all collaborative experiments are identified by the location  
of the laboratory where they were performed. Later claims for evidence by the MSU 
group  \cite{MSU} of non-negligible E2 contribution 
in inclusive measurement of an asymmetry, were not confirmed in 
a recent exclusive measurement of a similar asymmetry by the GSI2 
collaboration \cite{Sch03}.

It was then a great surprise to learn in a recent Physical Review Letter 
by Esbensen, Bertsch and Snover \cite{PRL} that higher order terms and 
an E2 contribution are an important correction to the RIKEN2 data \cite{PRL}.
They quote in their PRL paper \cite{PRL} statements from the Seattle group 
\cite{Jung03} and indeed the PRL paper is based on these claims \cite{Jung03}.
We demonstrate that these claims are based on misrepresentation of data and a
misunderstanding of the CD method.

\begin{figure}
\includegraphics[width=5in]{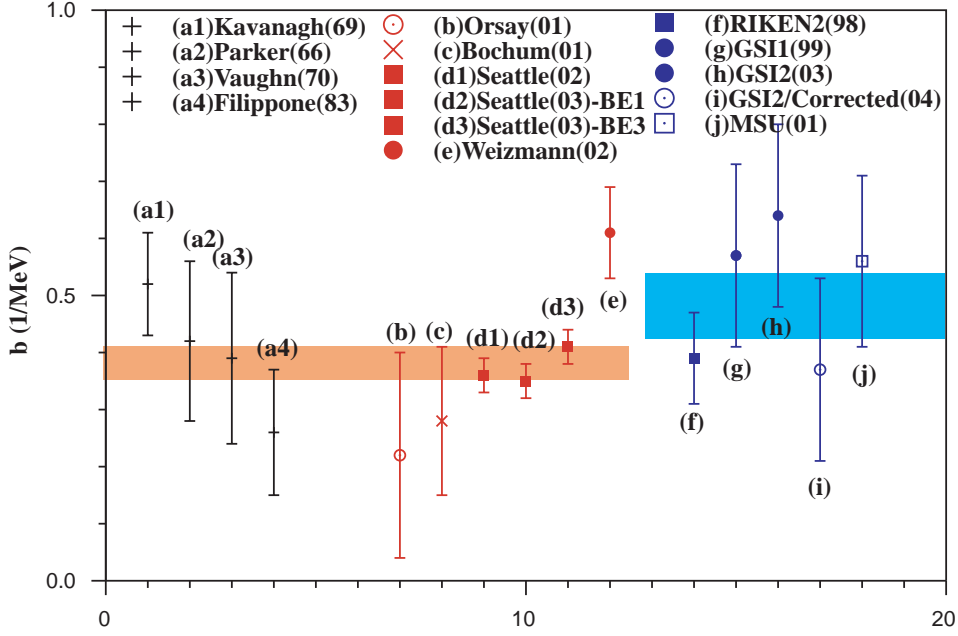}
 \caption{\label{Slopeb} The extracted so called "scale independent slopes (b)" 
of world data. We correctly plot the RIKEN2 data and add several data points 
that were neglected by the Seattle group \cite{Jung03} as discussed in the text. 
The range of "average values" is indicated and discussed in the text.}
\end{figure} 

\section{Misrepresentation of data on b-slope parameter}

The Seattle group chose to parameterize the astrophysical cross section 
factor, \xs17, as S = a(1+bE), and they show  in Fig. 19 of their paper \cite{Jung03}
the so called "scale independent slopes (b)" extracted for energies smaller 
than 425 keV and between 830 - 1300 keV. We first note that excluding 
the region of the 632 keV resonance is possibly relevant (but not necessary) 
for Direct Capture (DC) data, but in CD data the contribution of 
that state is reduced (by a factor of $\beta ^2$) and is 
negligible. For the RIKEN kinematics it is in fact smaller than the quoted error bar. 
The GSI CD data are quoted with that small contribution already removed and there 
is no reason to exclude a certain energy region to determine the slope of CD data. 

In Fig. 1 we show the b-slope parameter extracted from both DC and CD data in the 
energy range of 300 - 1400 keV. For the DC data we exclude the same energy region 
 \cite{Jung03} due to 632 keV resonance and we subtract its M1 contributions at higher 
energies. All so obtained fits have an acceptable \xchi \ close to unity. 
The obtained b-slope parameters agree with Ref. \cite{Jung03}, however the 
so called scale independent b-slope parameter of the RIKEN2 data \cite{Kik} was 
plotted incorrectly by the Seattle group in Fig. 19 of their paper \cite{Jung03}. It
is plotted correctly in Fig. 1, where we also include three DC data points that 
were omitted by the Seattle group \cite{Jung03}, as we discuss below.

A detailed discussion of 
the difference between the RIKEN2 and GSI1 data was already presented in the literature 
\cite{Sch03}. For example from Fig. 4 of the GSI2 paper \cite{Sch03} it is clear that the 
central value of the slope of the RIKEN2 data is smaller than the central value of the GSI1 
data \cite{Iw99}, but they are plotted by  the Seattle group in Fig. 19 with almost the same 
central value. We can not explain the origin of this simple mistake of the Seattle group 
\cite{Priv}, which in of itself might be of no consequence, except for  
its relevance for the newly published PRL paper \cite{PRL} as we discuss 
below.

Very recently it was also suggested \cite{Klaus,NIC8} that a reanalysis 
of the GSI2 data may yield a smaller slope for the GSI2 data, as also shown in Fig. 1.
This latest analysis has yet to be scrutinized \cite{Sch03} and 
published, none-the-less it is shown here since it was already presented in international 
meetings \cite{Klaus,NIC8}. But this result  \cite{Klaus,NIC8}, as well as all other 
non published results, are not included in the calculation of the average 
slope and extrapolated \S17 values of CD data. 

The published  RIKEN2 \cite{Kik}, GSI1\cite{Iw99}, GSI2 \cite{Sch03}, 
and MSU data \cite{MSU}, yield a 1/$\sigma$ weighted average for the 
"b-slope parameter", b = 0.51 \xpm 0.06 $MeV^{-1}$
with \xchi = 1.1. With a central value that is slightly smaller than quoted in 
Ref. \cite{Jung03}, b = 0.55 \xpm 0.08  and \xchi = 0.2. The 
quoted small \xchi \ is due to the wrong value used for the RIKEN2 data 
in Ref. \cite{Jung03}.
We also note with much disappointment that the discussed average b-slope 
parameters \cite{Jung03} are a factor of 10 larger than the very same b-slope 
parameters shown in Fig. 19 \cite{Jung03}. The smaller values plotted 
in Fig. 19 are the correct ones.

In the same figure the Seattle group suggests that a remarkable agreement exists between 
measured b-slope parameters of DC data \cite{Jung03,Ham01,Str01,Weiz,Fil83}.
The DC data shown in Fig. 19 \cite{Jung03} does not include all available data.
We note that the "old" data of Filippone \cite{Fil83} are included \cite{Jung03},
but the data measured by  Vaughn {\em et al.} \cite{Vaughn}, Parker \cite{Parker}, and 
Kavangh {\em et al.} \cite{Kav} were ignored in their Fig. 19. The last three data sets  
were deemed \cite{Adel} not useful at low energies for extrapolating \S17, but they are certainly 
useful for studying the slope of the data measured at energies between 300 and 1400 keV as done 
in Fig. 19 of the Seattle group. Most disturbing is the fact that they exclude the published data 
of Parker \cite{Parker} and Vaughn {\em et al.} \cite{Vaughn} but included that of 
Filippone \cite{Fil83}. As we show below including these data as well as including the 
correct value of the RIKEN2 data reduces the claimed disagreement by a factor 2.

\section{Disagreement Among Direct Capture Data}

It is well known that the "old" data on DC \be7pg reaction \cite{Fil83,Vaughn,Parker,Kav} 
exhibit major systematic disagreements. But the situation 
is not improved with "modern" data on DC
\cite{Jung03,Ham01,Str01,Weiz}. The data of the Orsay group \cite{Ham01} 
and Bochum group \cite{Str01} do not agree with that of the Seattle group 
\cite{Jung03} and Weizmann group \cite{Weiz}. The disagreement is by as much 
as five sigma and there is not a single measured data point of the Bochum group 
that agrees with a data point measured by the Seattle group. Most disturbing is the 
disagreement of the b-slope parameter extracted from the two "modern" high precision
measurements of the Seattle \cite{Jung03} and Weizmann \cite{Weiz} groups, 
shown in Fig. 19 of \cite{Jung03} as well as in Fig. 1 of this paper. 
From Fig. 19 of \cite{Jung03} it is clear 
that the b-slope parameter of the Orsay \cite{Ham01}, Seattle \cite{Jung03} and Weizmann 
\cite{Weiz} data do not have overlapping error bars. As we show below the, disagreement between 
the slope parameters is in fact more significant. The slightly better agreement between 
"modern" DC data is an artifact of the parametrization used by the Seattle group.

The large systematical disagreement between DC data 
can not be handled algebraically to extract a meaningful 
average slope. For example the slopes of the Orsay \cite{Ham01}, Seattle 
\cite{Jung03} and Weizmann \cite{Weiz} shown in Fig. 19 \cite{Jung03} can not 
be all true if the error bars shown in Fig. 19 \cite{Jung03} are correct. 
None-the-less it has been customary to artificially enlarge the 
error bars by multiplying it by the square root of \xchi, so as to make data with 
systematical differences appear as if it is statistically distributed. Using such 
a procedure for all published DC data 
\cite{Ham01,Str01,Weiz,Fil83,Vaughn,Parker} and the BE1 and BE3 data sets 
of the Seattle group \cite{Jung03} we extract a $1/\sigma$ weighted average 
b-slope parameter, b = 0.38 \xpm 0.02 
$MeV^{-1}$ with \xchi = 1.8. Taking into account the bad \xchi \, as discussed above, 
 we obtain b = 0.38 \xpm 0.03 $MeV^{-1}$
 as compared to b = 0.311 \xpm 0.014 $MeV^{-1}$ and \xchi = 1.9, quoted by the 
 Seattle group. 
 
 We conclude that the b-slope parameter can not be extracted
 from DC data with the (impressive) accuracy of 4.5\% \cite{Jung03}, 
 unless one excludes some of the DC 
 measurements discussed above. An error which is approximately a factor of 
 2 larger seems like a more reasonable choice. Also a difference of the 
 extracted b-slope parameter of 0.25 $MeV^{-1}$ quoted by the Seattle group \cite{Jung03} is 
 not confirmed in this analysis where we find a difference which is approximately 
 a factor 2 smaller.
 
 The so obtained average b-slope parameters in CD, b = 0.51 \xpm 0.06 $MeV^{-1}$, differs  
by only 1.9 sigma from that extracted for DC data,  b = 0.38 \xpm 0.03 
 $MeV^{-1}$, and can hardly be characterized as significant. This slight difference 
 is particularly negligible in view of the substantial systematical disagreement among 
 individual data sets measured in DC experiments. Furthermore, it should not be 
 considered sufficient to motivate new theoretical 
 investigation or for that matter a publication in the prestigious 
 Journal of the Physical Review Letters. As we show below the actual differences 
 of measured slopes are even smaller.

\section{Lack of Physical Justification of the Seattle Slope Parameterization}

We now demonstrate that the parametrization S = a(1 + bE) used by the Seattle 
group \cite{Jung03} has no physical justification.
We refer the reader to Baye's seminal paper \cite{Baye} on the theory of DC 
where we find that for an external capture reaction (and only when the conditions for an 
external capture reaction are satisfied, i.e. below 100 keV):

\begin{tabbing}

\hspace{2in}  \= $S(E)$  \=  = $S(0)[1 \ + \  s_{1} \times E]$   \hspace{0.3in}  \= (equ 1) \\
\     \\
          \> $S(0)$  \>  = $S_d(0) + \  S_s(0)$  \> (equ 2)  \\
          \     \\
                  \> and, \\
              
         \> $s_1$ \>  = ${S_s(0) \over S(0)} [s_{1s} \ + s_{1d}  \times {S_d(0) \over S_s(0)}]$ \> (equ 3)\\
    
\end{tabbing}

Equation 1-3 were derived for external capture and 
are correct only when the conditions for external capture are met.
Specifically the value of $s_1$ is explicitly negative as predicted 
for energies below 100 keV. 
In such a case the logarithmic derivative S'/S(0) is shown to be 
an invariant \cite{Baye}. 

At energies above 300 keV 
the observed slope are manifestly positive. For higher energies 
one must add higher orders beyond the linear term used in equ 1, 
so as to turn the sign of the slope from negative to positive.
For example the Taylor expansion of the theoretical curve predicted 
by Descouvemont and Baye \cite{DB} can not be truncated 
below third order to yield a reasonable representation of the predicted curve
at all energies up to 1.5 MeV. Clearly 
at energies above 300 keV the truncation of the Taylor expansion to a linear 
term, leads to unphysical expansion.

Furthermore, as we show in Fig. 2 the slope of the d-wave component is essentially constant 
as a function of energy \cite{Jen} (approximately +10 eV-b/MeV), but the slope of the s-wave 
component varies with energy between approximately -20 eV-b/MeV at zero energy 
to approximately -3 eV-b/MeV at higher energies.  The overall slope 
(S' = dS/dE) is a sum of the two components which is energy dependent. At very 
low energies the s-wave negative slope is dominant and is almost constant hence the linear 
truncation as in equ. 1. At energies above 500 keV the d-wave slope dominates and the 
variation of the s-wave slope is small leading to almost constant positive 
slope. At high energies the linear dependent of the S-factor with a positive slope 
is an artifact of the d/s ratio. But the over all slope at higher energies is dependent on the 
low energy cut of the data. In particular the fit parameters "a" used in \cite{Jung03} is most 
sensitive to the choice of the low energy cutoff of the data between 300 and 500 keV, and 
the so-called scale independent b-slope parameter varies due to the selected range of 
data and not due to the physical slope (S').

In equ. 1 the over all normalization factor S(0) is directly related  
to the astrophysical cross section factor at zero energy, 
as well as the ANC of the physical wave function \cite{Baye}. But the fit parameter 
"a" used by the Seattle group \cite{Jung03}  
has no physical meaning and is not directly related to the ANC or the spectroscopic factor. 

Furthermore, the observed positive slope (S' = dS/dE) of the cross section factor at 
energies above 500 keV is directly related to the (model dependent) 
d-wave contribution designated by d/s ratio \cite{Fil83,Baye,Jen,Hamish,Barker,Xu94}. 
It is self evident that the d/s ratio is independent of S(0), since both d 
and s components are proportional to the same ANC \cite{Baye}. 
Hence it is clear that at energies 
above 300 keV the slope (S') is {\bf not} directly related to the over all 
normalization S(0) and the so called "scale independent slope (b)" 
defined by the Seattle group \cite{Jung03} has no physical justification.

\begin{figure}
\includegraphics[width=5in]{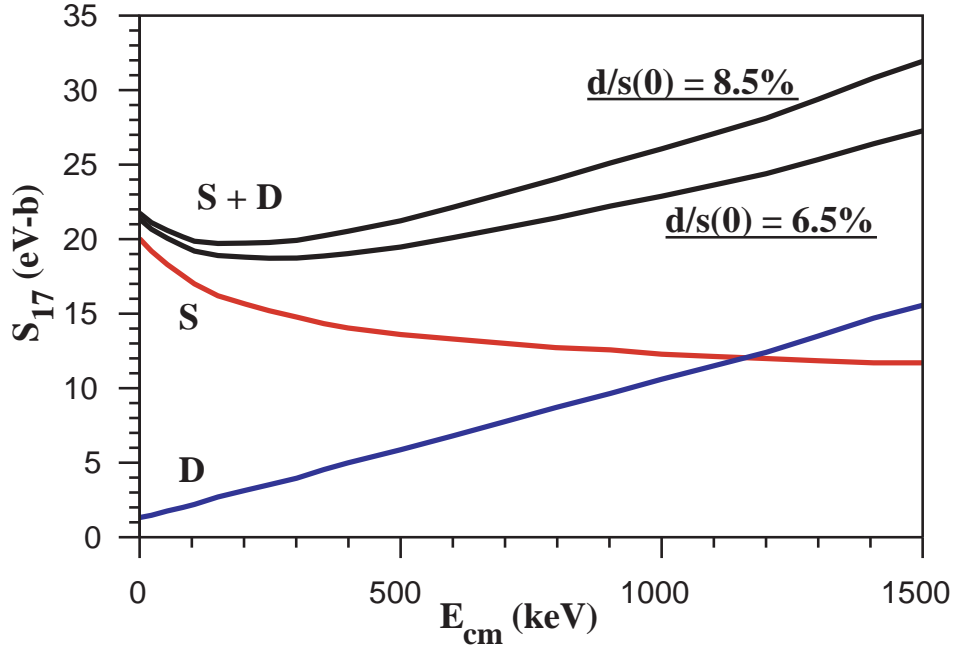}
 \caption{\label{Jennings} The s and d wave contributions calculated by 
Jennings {\em et al.} \cite{Jen} and with an increased d/s(0) = 8.5\% value.}
\end{figure} 

To illustrate this point we show in Fig. 2 the s and d waves contributions calculated by 
Jennings {\em et al.} \cite{Jen}. In the same figure we show the predicted 
S-factor with $S_d (0)$ that is increased  
from 1.3 eV-b \cite{Jen} to 1.7 eV-b, which is certainly within the limit of 
accuracy of theoretical predictions. This yield a very insignificant (1.8\%) 
change of S(0) but a very significant change ($\approx$ 25\%) of the 
observed slope (S') above 300 keV. This schematic model most vividly demonstrates 
the fact that the parametrization S = a(1+bE) has no physical justification 
nor does the assumption that S'/S(0) above 300 keV is an invariant.

We add that a knowledge of the d-wave component at measured energies 
(e.g. above 300 keV) is essential for extrapolating \S17.
For example at 500 keV the d-wave contribution amounts to 30\% of the 
measured \xs17 and above 1,150 keV it is dominant. 
At zero energy, on the other hand, the d-wave contribution is predicted to be 
small ($S_d(0)/S_s(0) \ \approx \ 6\%$).
Thus in order to accurately extrapolate \xs17 to zero energies one must remove 
the d-wave contribution from measured \xs17, as has been emphasized long ago 
by Robertson \cite{Hamish}. This so far has been done 
by means of theoretical estimate of the d-wave component and a chi-square 
fit of data by the predicted s + d wave components. 

But currently there is no 
direct way to test the validity of the model dependent prediction of the d-wave 
component. In fact the slope (S') seems thus 
far the only way to determine the d-wave contribution. As we show below the 
slope has not been measured with high accuracy and large discrepancies 
still exist between measured DC data. Indeed 
the ill defined theoretical d/s ratio leads to an uncertainty of the extrapolation. 
Other theoretical issues were also discussed by Descouvemnot and they yield 
an uncertainty of at least 6\% \cite{D}, a factor of 2 larger than the theoretical 
uncertainty quoted by in Ref. \cite{Jung03}.

\section{The Slope S'}

\begin{figure}
\includegraphics[width=5in]{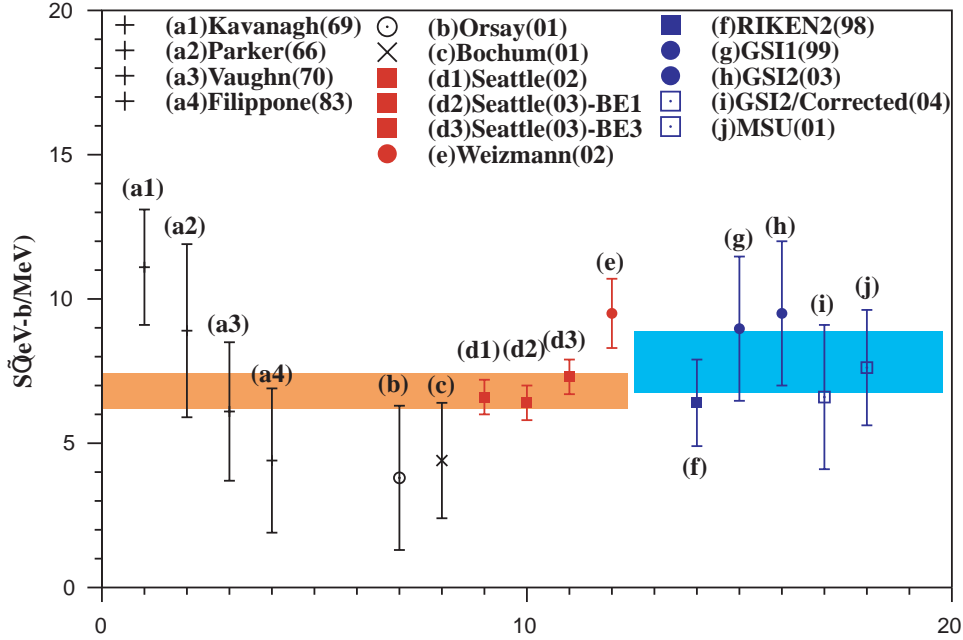}
 \caption{\label{SlopeS} The measured slopes (S' = dS/dE) 
of world data measured between 300 and 1400 keV, as discussed in 
the text. The range of "average values" is indicated and discussed in the text.}
\end{figure} 

For that reason we plot in Fig. 3 the extracted slope S' = dS/dE for data 
measured between 300 and 1400 keV, with the same exclusion of DC data 
due to 632 keV resonance between 450 and 850 keV, and correcting for the 
M1 contribution of that resonance above 850 KeV as discussed in section II.
Figure 3 includes all published data and the slope of the RIKEN2 data is 
plotted correctly.  In this figure we note a sharper disagreement between 
"modern" DC data. In particular we extract three different slopes values for  
the four "modern" experiments on DC:  a Bochum-Orsay low value \cite{Ham01,Str01}, 
an  average of the three Seattle results \cite{Jung03} and the Weizmann higher value.
\cite{Weiz}.  These three values of "modern" DC  slopes do not 
exhibit overlapping error bars. In sharp contrast all four published 
CD slopes shown in Fig. 3 are in agreement within the quoted error bars. 

From these data we extract for CD data the 1/$\sigma$ 
weighted average S' = 7.9 \xpm 1.0 
eV-b/MeV with \xchi = 0.5 and for DC data S' = 6.7 \xpm 0.5 
eV-b/MeV with \xchi = 1.5, hence S' = 6.7 \xpm 0.6 eV-b/MeV. Clearly the 
average slope extracted in CD data agrees (within 1.0 sigma) with that extracted from  
DC data. This agreement is considerably better than observed 
between individual DC measurements, as we discuss above.

We conclude that there is no significant disagreement between slopes 
extracted from CD and DC data. This removes the very raison d'etre 
for the recent PRL paper \cite{PRL}. 

\section{Comparison of RIKEN2 Data with Theory of Esbensen, Bertsch and Snover}

Furthermore, in this PRL paper \cite{PRL} we 
find a substantial (50\%) correction of the b-slope parameter extracted for the RIKEN2 
data that is implied to be 0.25 $MeV^{-1}$.
As we discuss above the b-slope parameter was computed incorrectly 
by the Seattle group, as shown in their Fig. 19 \cite{Jung03}.
For the RIKEN2 data b = 0.39 \xpm 0.08  $MeV^{-1}$, and the said correction 
yields: b = 0.14 \xpm 0.08 $MeV^{-1}$ which is considerably (more than 
a factor of 2) smaller than the observed central value for DC data, 
b = 0.38 \xpm 0.03 $MeV^{-1}$. We conclude that the corrections suggested 
by Esbensen, Bertsch and Snover, in sharp contrast to the claim \cite{PRL}, in fact 
lead to a disagreement between the RIKEN2 data and DC data, and are not relevant 
at least for the RIKEN2 data, and in particular they once again confirm that the E2 
correction is negligible.

\section{\S17 Extracted From CD Data}

In Fig. 20 of the Seattle paper \cite{Jung03} they show extracted \S17 from CD
using the extrapolation procedure of Descouvemont and Baye \cite{DB}, 
and based on this analysis it is stated \cite{PRL} that "the zero-energy extrapolated 
\S17 values inferred from CD measurements are, on the average 10\% 
lower than the mean of modern direct measurements". The extracted 
\S17 shown in Fig. 20 \cite{Jung03} are only from data measured at energies below 
425 keV and the majority of CD data points that were measured above 425 keV 
were excluded in Fig. 20 \cite{Jung03}. 

This arbitrary exclusion of (CD) data above 
425 keV has no physical justification. For example 
as shown by Descouvemont \cite{D} the theoretical error increases to 
approximately 5\% at 500 keV and in fact it is slightly decreased up to 
approximately 1.0 MeV, and there is no theoretical justification for 
excluding data between 500 keV and 1.0 MeV. 
In DC measurements the well known contribution of the 632 keV 
resonance needs to be subtracted, but that is not the case for example 
in RIKEN CD data.
Furthermore, as we discussed above, the slope of the 
data between 300 and 1,400 keV is essential for determining the d/s 
ratio and the extrapolation to zero energy. Excluding data above 425 keV 
reduces our sensitivity for testing the various model prediction of the d/s ratio.

When including CD data measured only at energies below 425 one runs 
into a more serious systematical problem. Namely, the 
relative energy measured in CD ($E_{rel}$) 
is determined mostly from the proton-$^7Be$ relative angle ($\theta_{17}$). At  
small relative energies (as well as at small scattering angles, $\theta_8$) plural  
scattering in the target are of major concern. These are estimated 
theoretically and are known to be inaccurate. The effect of plural scattering in 
the target is indeed known for practitioners in the field of CD and has been discussed 
on several occasions. It leads to a systematical uncertainty of the measured 
\xs17 of the order of 2 eV-b.

Furthermore, the yield of the CD of \b8 arises from a convolution of the nuclear 
cross section which is rapidly dropping toward low energies, and the virtual 
photon flux that is rapidly increasing toward low energies. This generate a yield 
that is almost constant (\xpm 20\%) at $E_{rel}$ = 300 - 800 keV. Note that over 
the same energy range the DC yield changes by almost a factor of 10. 

Thus when excluding the CD data above 425 keV, the Seattle group excluded the 
data that were measured with the best accuracy and with smallest systematical 
uncertainty. This exclusion of "the best energy region" measured in CD experiment 
is a manifestation of their misunderstanding of the CD method. If in 
fact one insists on such a flawed analysis of CD data, one must estimate the 
systematic uncertainty due to this selection of data. This has not been done in 
the Seattle re-analyses of CD data \cite{Jung03}.

Instead we rely here on the original analyses of the authors.
In Fig. 4 we show the \S17 factors extracted by the original authors who 
performed the CD experiments. These results include all measured data points, 
and are consistently analyzed with the extrapolation procedure of Descouvemont 
and Baye \cite{DB}. The potential model of Typel \cite{Sch03} was also used 
in the GSI2 paper, but the so quoted \S17 are not shown in Fig. 4.

\begin{figure}
\includegraphics[width=5in]{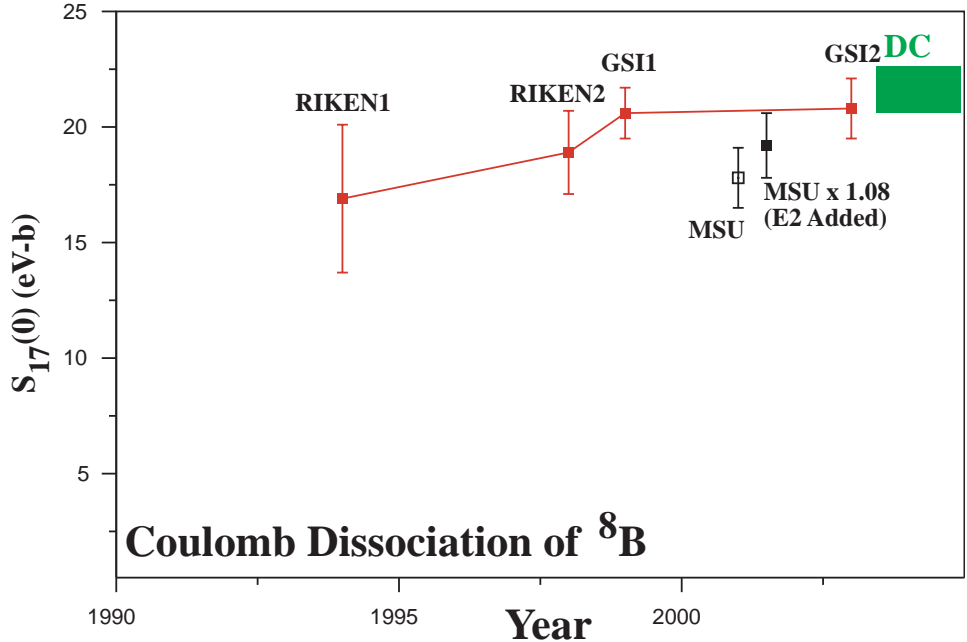}
 \caption{\label{CDDB} Measured \S17 as originally published by the authors 
who performed the CD experiments. These analyses include all measured data 
points \cite{Mot94,Kik,Iw99,Sch03,MSU} using the 
extrapolation procedure of Descouvemont and Baye \cite{DB}. We also plot 
the MSU data with the E2 correction ($\approx 8\%$) \cite{MSU} added back 
to the quoted \S17, as  discussed in the text. The range of \S17 results from the 
measurements of DC by the Seattle \cite{Jung03} and Weizmann groups 
\cite{Weiz} is indicated.}
\end{figure} 

We note that the five experiments on the CD of \b8 \cite{Mot94,Kik,Iw99,Sch03,MSU} 
show a remarkably good agreement within the quoted error bars, in sharp contrast to 
the confusion that exists in "old" \cite{Fil83,Vaughn,Parker,Kav} and 
"modern" DC results \cite{Jung03,Ham01,Str01,Weiz}. The results 
of the RIKEN-GSI experiments must be considered as a continuation of the 
same experiment (essentially the same experiment repeated four times) 
with an improved kinematical and experimental conditions. 
Thus the four results can not all be assigned the same weight. 

We note that while 
these four results are consistent within the quoted error bar, they show a systematic 
trend of an increased \S17 (to approximately 20.7 eV-b), while the error 
bars are reduced. The MSU result on the other hand 
includes a model dependent E2 correction ($\approx 8\%$) deduced from inclusive 
experiments \cite{MSU}, which was not confirmed in a recent exclusive measurement  
of a similar asymmetry \cite{Sch03}. When this E2 correction is added back to 
the quoted MSU result \cite{MSU}, as shown in Fig. 4, together with the published 
RIKEN2 \cite{Kik}, GSI1 \cite{Iw99}, and GSI2 \cite{Sch03} results we obtain 
a 1/$\sigma$ weighted average of \S17 = 20.0 \xpm 0.7 with \xchi = 0.5, 
which is in excellent agreement with the measurement of the Weizmann group 
\cite{Weiz} and in agreement with the measurement of the Seattle group \cite{Jung03}.

We conclude that the statements (which were repeated in the literature) on substantial 
differences between \xs17  data from DC and CD measurements as well as extracted \S17 
are  based on misrepresentation of data, misunderstanding of theory and a biassed 
approach of the Seattle group \cite{Jung03}. Quite to the contrary we find a good 
agreement among CD measurements as well as an agreement of the CD 
results with the two most recent high precision results of DC experiments. Further 
attention must be given to an accurate measurement of the slope (S') and the 
d-wave contribution to \xs17 measured at energies above 300 keV so as to allow 
accurate extrapolation to zero energy.


\begin{thebibliography}{9}

\bibitem{Baur} G. Baur, C.A. Bertulani, and H. Rebel;
 Nucl. Phys. {\bf A458}(1986)188.

\bibitem{Mot94} T. Motobayashi {\em et al.}; Phys. Rev. Lett. {\bf 73}(1994)2680.

\bibitem{Kik} T. Kikuchi {\em et al.}; Phys. Lett. {\bf B391}(1997)261, 
 ibid E. Phys. J. {\bf A3}(1998)213.

\bibitem{Iw99} N. Iwasa {\em et al.}; Phys. Rev. Lett. {\bf 83}(1999)2910.

\bibitem{Sch03} F. Schumann {\em et al.}; Phys. Rev. Lett. {\bf 90}(2003)232501.

\bibitem{Gai} M. Gai and C.A. Bertulani; Phys. Rev. {\bf C52}(1995)1706.

\bibitem{MSU} B.S. Davids {\em et al.}; Phy. Rev. {\bf 63}(2001)065806.

\bibitem{PRL} H. Esbensen, G.F. Bertsch, and K. Snover; Phys. Rev. Lett. 
   {\bf 94}(2005)042502.

\bibitem{Jung03} A.R. Junghans {\em et al.}; Phys. Rev. {\bf C68}(2003)065803.

\bibitem{Klaus} K. Suemmerer, contribution, Advances and Challenges in Nuclear 
Astrophysics, ECT* Workshop, May 24-28, 2004, Trento.
 

\bibitem{NIC8} F. Schumann for the GSI S223 Experiment, contribution, Nuclei in the 
 Cosmos 8, Vancouver, July 19-24, 2004.
 
 \bibitem{Priv} Requests  including a copy of Fig. 4 of \cite{Sch03} sent to the lead 
 author (Dr. Snover) to correct this mistake were ignored, as was a request to receive the 
 numerical values of data and the resultant "a"  and "b" fit parameters. 
 A recent linear fit by Dr. Bertsch of the RIKEN2 data 
 yields b-slope parameter that is considerably smaller than plotted in  
 Fig. 19 of \cite{Jung03}.
 
\bibitem{Ham01} F. Hammache {\em et al.}; Phys. Rev. Lett. {\bf 86}(2001)3985.

\bibitem{Str01} F. Strieder {\em et al.}; Nucl. Phys. {\bf A696}(2001)219.

\bibitem{Weiz} L.T. Baby Phys. Rev. {\bf C67}(2003)065805, ER {\bf C69}(2004)019902(E).

\bibitem{Fil83} B.W. Filippone {\em et al.}; Phys. Rev. {\bf C28}(1983)2222.

\bibitem{Vaughn} F.J. Vaughn {\em et al.}, Phys. Rev. {\bf C2}(1970)1657.

\bibitem{Parker} P.D. Parker, Phys. Rev. {\bf 150}(1966)851.

\bibitem{Kav} R.W. Kavanagh, T.A. Tombrello, T.A. Mosher, and 
 D.R. Goosman, Bull. Amer. Phys. Soc., {\bf 14}(1969)1209.

\bibitem{Adel} E.G. Adelberger {\em et al.}; Rev. of Modern Phys. {\bf 70}(1998)1265.

\bibitem{Baye} D. Baye; Phys. Rev. {\bf C62}(2000)065803.

\bibitem{DB} P. Descouvemont and D. Baye; Nucl. Phys. {\bf A567}(1994)341.

\bibitem{Jen} B.K. Jennings, S. Karataglidis, and T.D. Shoppa; Phys. Rev. 
 {\bf C58}(1998)3711. 

\bibitem{Hamish} R.G.H. Robertson; Phys. Rev. {\bf C7}(1973)543.

\bibitem{Barker} F.C. Barker; Aust. Jour. Phys. {\bf 33}(1980)177.
\bibitem{Xu94} H.M. Xu {\em et al.}; Phys. Rev. Lett. {\bf 73}(1994)2027.

\bibitem{D} P. Descouvemont; Phys. Rev. {\bf C70}(2004)065802.

\end{thebibliography}
\end{document}